\documentclass[reprint, preprintnumbers, superscriptaddress, nofootinbib, amsmath, amssymb, aps, prd, floatfix]{revtex4-2}
\usepackage{mathtools}
\usepackage{lineno,hyperref}
\usepackage{orcidlink}
\hypersetup{colorlinks=true}
\usepackage{color, units}
\usepackage{appendix}
\usepackage{comment}
\usepackage[T1]{fontenc}
\usepackage[normalem]{ulem} % Strike out
\hypersetup{
 colorlinks=true,
 linktoc=all,
 linkcolor=blue,
 citecolor=blue,
 urlcolor = blue}

\def\tdfstat{TDFstat}
\newcommand{\fstat}{{$\mathcal{F}$-statistic}}
\newcommand{\dos}{D_\text{S}}
\newcommand{\dol}{D_\text{L}}
\newcommand{\dls}{D_\text{LS}}
\newcommand{\dr}{D_\text{r}}
\newcommand{\vrel}{V_\text{rel}}

\newcommand{\w}{W}
\newcommand{\ml}{M_\text{L}}
\newcommand{\te}{T_\text{E}}
\newcommand{\re}{R_\text{E}}
\newcommand{\tobs}{T_\text{obs}}
\newcommand{\ts}{T_\text{s}}

\newcommand{\bs}[1]{\boldsymbol{#1}}
\newcommand{\CAMK}{Nicolaus Copernicus Astronomical Center, Polish Academy of Sciences, Bartycka 18, 00-716, Warsaw, Poland}
\newcommand{\INFN}{INFN Sezione di Ferrara, Via Saragat 1, 44122 Ferrara, Italy }
\newcommand{\NCBJ}{National Centre for Nuclear Research, Pasteura 7, 02-093 Warsaw, Poland}
\begin{document}
\title{Microlensing of long-duration gravitational wave signals originating from Galactic sources}

\author{Sudhagar Suyamprakasam $^*$\orcidlink{0000-0001-8578-4665}}
\email{Corresponding author: sudhagar@camk.edu.pl}
\affiliation{\CAMK} 

\author{Sreekanth Harikumar\orcidlink{0000-0002-2653-7282}}
\email{sreekanth@camk.edu.pl}
\affiliation{\CAMK}
\affiliation{\NCBJ}

\author{Paweł Ciecieląg\orcidlink{0000-0002-5871-4730}}
\affiliation{\CAMK}

\author{Przemysław Figura\orcidlink{0000-0002-8925-0393}}
\affiliation{\CAMK}

\author{Michał Bejger\orcidlink{0000-0002-4991-8213}}
\email{bejger@camk.edu.pl}
\affiliation{\CAMK}
\affiliation{\INFN}

\author{Marek Biesiada\orcidlink{0000-0003-1308-7304}}
\email{marek.biesiada@ncbj.gov.pl}
\affiliation{\NCBJ}
%\date{\today}
\begin{abstract}
Detection of quasi-monochromatic, long-duration (continuous) gravitational wave radiation emitted by, e.g., asymmetric rotating neutron stars in our Galaxy requires a long observation time to distinguish it from the detector’s noise. If this signal is additionally microlensed by a lensing object located in the Galaxy, its magnitude would be temporarily magnified, which may lead to its discovery and allow probing of the physical nature of the lensing object and the source. We study the observational effect of microlensing of continuous gravitational wave signals for Galactic sources and lenses in the point mass lens approximation. In particular, we examine the regions of the parameter space that are promising for lensed CW searches, and perform example simulations to demonstrate how the lensing effect affects the continuous-wave signal. We show that an analytical lensing pattern can be identified from the lensed continuous wave signal using the Time-Domain F-statistic search, as the estimated signal-to-noise ratio in each time-domain segment scales directly with the amplification factor.
\end{abstract}
\maketitle
% ===================================================== %
% ===================================================== %
\section{Introduction}
To date, the LIGO-Virgo-KAGRA (LVK) collaboration, operating a global network of ground-based gravitational-wave (GW) detectors \cite{LIGO_2015, Virgo_2014, KAGRA_2019} has conducted several searches for quasi-monochromatic, continuous gravitational wave (CW) signals by examining all-sky directions and frequencies without prior information about the source's position and frequency parameters~\cite{PhysRevD.77.022001, PhysRevD.85.022001, Aasi_201408, PhysRevD.90.062010, PhysRevD.105.122001, PhysRevD.106.102008}, assuming that the sky position of the source is known~\cite{Aasi_2015, PhysRevD.95.082005, PhysRevD.105.082005, PhysRevD.106.042003} or assuming that both sky position and frequency parameters are known~\cite{Abbott_2010, Aasi_201404, Abbott_2017, PhysRevD.99.122002, Abbott_202206}. Potential sources of these signals include asymmetric rotating neutron stars (NS); another possible source of CW is ultralight scalar bosons surrounding spinning black holes (BH)~\cite{PhysRevD.105.102001}, or light - planetary or asteroid mass - primordial BH (PBH) binaries during their in-spiral phase. For recent reviews on astrophysics of CW, refer ~\cite{Riles2023, Haskell2023, WETTE2023102880}. Although no confident detection has been reported so far, ongoing efforts continue to improve the limits on astrophysical parameters.

Detecting CWs presents significant challenges both instrumentally and computationally due to their inherently smaller GW amplitudes than the short-duration transient GW emitted by compact binary coalescences (CBC) already detected~\cite{PhysRevX.9.031040, PhysRevX.11.021053, PhysRevD.109.022001, PhysRevX.13.041039}. Specifically, CW searches require long observing times to increase the signal-to-noise ratio (SNR), which grows as the square root of observing time~\cite{universe5110217}, and require substantial computational resources to analyze and post-process a large volume of data, especially for all-sky searches.

GW traveling along null geodesics may undergo strong gravitational lensing in analogy to similar phenomena known and observed in the electromagnetic (EM) domain~\cite{Gravitational_lenses1992,strong_weak_micro}. To date, the majority of studies have concentrated on the lensing effect in the context of short-duration transient GW signals~\cite{PhysRevLett.77.2875, PhysRevLett.80.1138, Takahashi_2003, 2017arXiv170204724D, PhysRevD.98.103022, universe9050200} from cosmologically distant sources, which allows multiple copies of the same signal to arrive at the detector at different times with altered amplitudes and phases. Detecting these signals would provide a range of applications, including cosmology~\cite{PhysRevLett.130.261401, NatureComm2017, SciRep2019}, tests of general relativity~\cite{Goyal:2020bkm,PPN2024, PRL2017, Ezquiaga:2020dao, Harikumar:2023gzh} and improved sky-localization~\cite{Wempe:2022zlk, 10.1093/mnras/staa2577}. Ongoing efforts aim to identify gravitational lensing signatures from these short-duration transient GW sources; however, no conclusive evidence of lensing has yet been reported~\cite{Abbott_2021b, Abbott_202407}. Predictions for future generations of GW detectors appear more optimistic concerning the detectability of strongly lensed CBC signals~\cite{Aleksandra2013, Ding_2015,DECIGO2021}. In the case of the Einstein Telescope, the expected rates could be as high as 50 to 100 lensed GW signals from CBCs per year~\cite{Lilan2022}. 

Most studies on the gravitational lensing of CWs have focused on two primary scenarios: strong lensing, which typically involves supermassive black hole (SMBH), such as the central BH in our Galaxy, serving as lensing objects~\cite{Ruffa_1999, Basak_2023, PhysRevD.109.024064, federico_muciaccia}, and microlensing, which examines star clusters as lensing objects~\cite{Moylan:2007fi, Suvorov_2022}. Some studies have explored the detectability of microlensing effects, including diffractive, interference, and beat patterns in CWs~\cite{Liao_2019, universe7120502}. There are several key differences between short-duration CBC signals and CWs. First, the most realistic sources of CWs are expected to be of Galactic origin. Second, unlike in the case of a short-duration transient signal, the frequency of the CW signal remains nearly constant. At the same time, the position of the source relative to the lens could change over the observation time, leading to a microlensing effect similar to that observed routinely in EM~\cite{Paczynski, Gravitational_lenses1992, strong_weak_micro}.

In this paper, we venture beyond the constant amplitude of a CW signal and study the temporal variation of the amplitude due to the microlensing effect. We also explore the parameter space to assess this microlensing impact on galactic sources that are lensed by galactic objects, particularly from the perspective of ground based detectors. Detecting lensed CW signals would enable us to gain insight into the properties of sources such as NS~\cite{Haskell2023} and PHB~\cite{Oguri:2020ldf,Diego:2019rzc}.

The article is organized as follows: In Sec.~\ref{sec:lensing_intro}, we provide a concise overview of gravitational lensing theory, emphasizing the differences in its application to short-duration transient signals and CW signals. We discuss the amplification factor for CW signals within the framework of point mass lens approximation and examine the lensing parameter space. Subsequently, we briefly describe the signals, outline the data preparation necessary for the simulations, and examine the Time Domain F-statistic ({\tdfstat}) analysis performed on both the unlensed and lensed signal in Sec.~\ref{sec:tdfstat}. Section~\ref{sec:discussion_conclusion} outlines possible future directions in the CW lensing field and summarizes our key findings.
% ===================================================== %
% SECTION 2 %
% ===================================================== %
\section{\label{sec:lensing_intro} Microlensing of Gravitational Waves}

\subsection{Basic concepts of gravitational lensing}
\begin{figure}[htp!]
    \centering
    \includegraphics[scale=0.5]{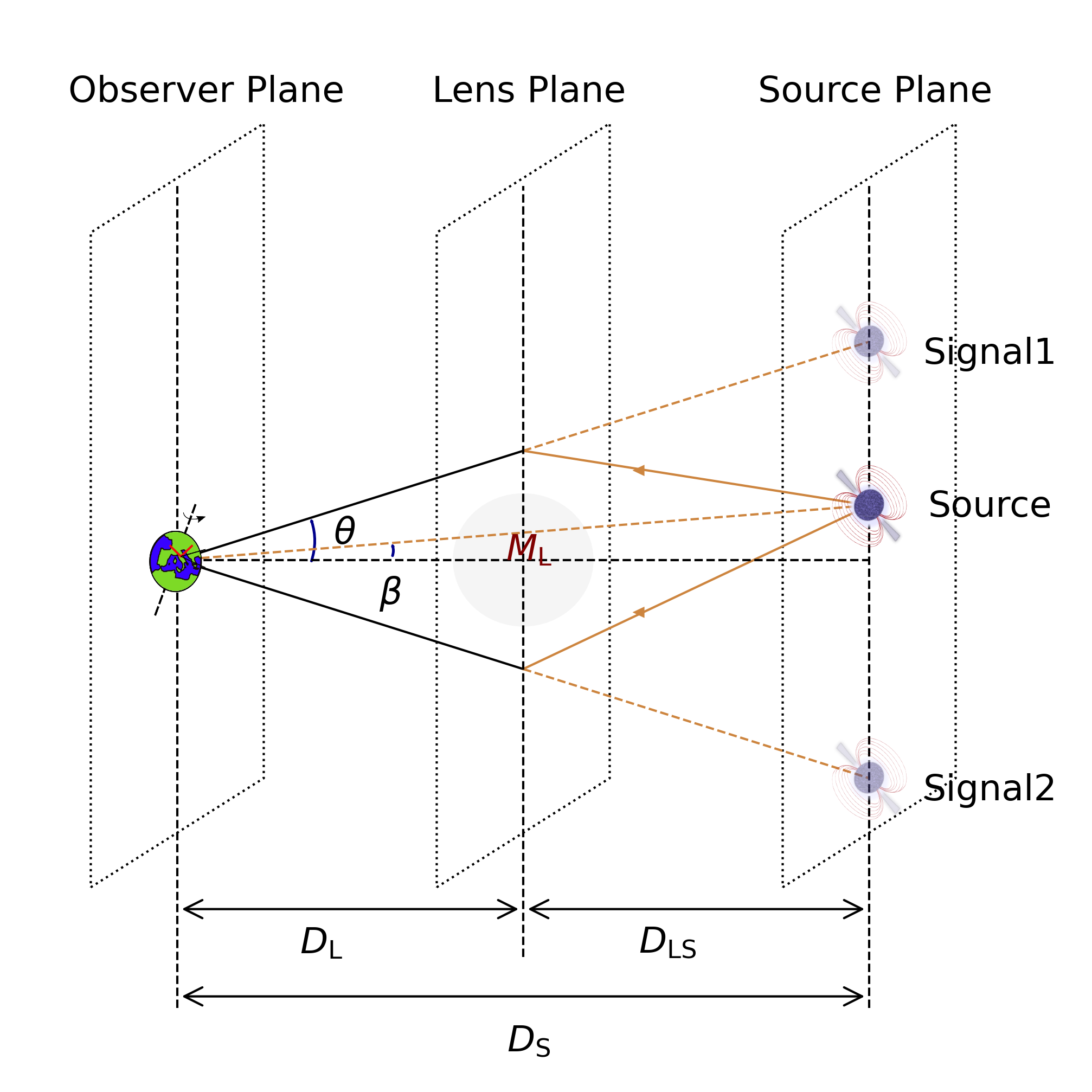}
    \caption{Strong lensing illustration. A lensing system consists of an observer, lens (of mass $M_L$), and source. The angle at which the observer sees the signal arriving at the lens plane is denoted as $\theta$, while the angle representing the position of the unlensed source as seen by the observer is referred to as $\beta$. $\dos$ represents the distance between the source and the observer, $\dls$ is the distance between the lens and the source. $\dol$ is the distance between the lens and the observer.}
    \label{fig:lensing-illust}
\end{figure}

In a gravitational lensing event, the signal from the source is influenced by a lensing object that lies along the line of sight between the source and the observer, as illustrated in Fig.~\ref{fig:lensing-illust}. An introduction to strong lensing theory can be found, for instance, in ~\cite{intro_to_lensing, Gravitational_lenses1992}, where much of the theory is presented within the framework of light-ray formalism in geometric optics. In our context, the wave optics regime is more appropriate~\cite{Oguri:2020ldf, Takahashi_2003}. Hence, we start our rudimentary introduction to this formalism. 
% ======================================== %
% Figure: Illustration Lensing System
% ======================================= %
In the wave optics formalism, the amplitude of GW obeys the Helmholtz equation, whose solution can be obtained from the Fresnel-Kirchhoff diffraction integral~\cite{Gravitational_lenses1992, universe9050200, universe7120502}. By solving the diffraction integral in the thin lens approximation, one can determine the effect of a lensing object on the incoming GW, expressed as the ratio of the lensed and unlensed GW amplitude~\cite{10.1143/PTPS.133.137, Gravitational_lenses1992, Takahashi_2003, Born:1999ory}. This term, also known as the amplification factor,\footnote{We use the symbol $\mathcal{A}$ to denote the amplification factor, which is a time-varying impact parameter at a fixed frequency $f$. In contrast, some authors use the symbol $F$ to represent frequency-dependent amplification at a constant impact parameter $y$.} has the following form
\begin{eqnarray} 
     \mathcal{A}(\w,\bs y) = \frac{\w}{2 \pi i} \int d^2 \bs x  e^{i \w T(\bs x, \bs y)}, 
     \label{eq:Awy}
\end{eqnarray}
where $\w$ is the dimensionless frequency defined as
\begin{equation}
    \w = \frac{2 \pi f \dos \xi^2_0 }{ c \dol \dls},
\end{equation}
where $f$ is the source's GW frequency, $\xi_0$ the characteristic length scale usually chosen as the Einstein radius $\re$ of the lens. The distances involved have their usual meaning, as shown in Fig.~\ref{fig:lensing-illust}, $\dls$ is the distance between the lens and the source, $\dos$, and $\dol$ are the distances to the source and lens measured from the observer, respectively. Let us introduce dimensionless quantities $\bs x$ and $\bs y$ as follows:
\begin{eqnarray}
    \bs x = \frac{\bs \xi}{\xi_0} \;\; ; \;\; \bs y = \frac{\bs \eta }{\eta_0},
\end{eqnarray}
where $ \bs\xi=\bs \theta D_{L}$ is the physical distance in the lens plane between the image and the center of the lens, $\bs \theta$ is the corresponding angular position of the image, $\bs \eta = \bs \beta  D_S$, and  $\bs \beta$  is the intrinsic angular position of the source, which the observer would measure in the absence of a lens. Similar to the characteristic length scale $\xi_0$ in the lens plane, there exists a corresponding length scale $\eta_0$ in the source plane related to $\xi_0$ and given by $\eta_0 = \xi_0 D_S/ D_L$. According to Fermat's principle, lensing affects the time of arrival of the GW. The total time delay induced by gravitational lensing at the position $\bs y$ of the lens plane  
is referred to as the time delay function or the Fermat potential~\cite{10.1143/PTPS.133.137, Gravitational_lenses1992}. It consists of two components: the geometric delay and the Shapiro time delay, and can be expressed as~\cite{10.1143/PTPS.133.137, Takahashi_2003}
\begin{equation}
    T(\bs x, \bs y) =  \frac{1}{2} | \bs x - \bs y|^2 - \psi(\bs x) + \phi_{m}(\bs y),
\end{equation}
where $\psi(x)$ is the effective lensing potential, which is the gravitational potential of the lens mass distribution projected onto the lens plane, and $\phi_{m}(y)$ is a constant picked to set the minimum value of the time-delay to zero. 

% ================================== %
% Figure: Amplification factor 2D
% ================================== %
\begin{figure}[htp!]
    \centering
	\includegraphics[scale=0.6]{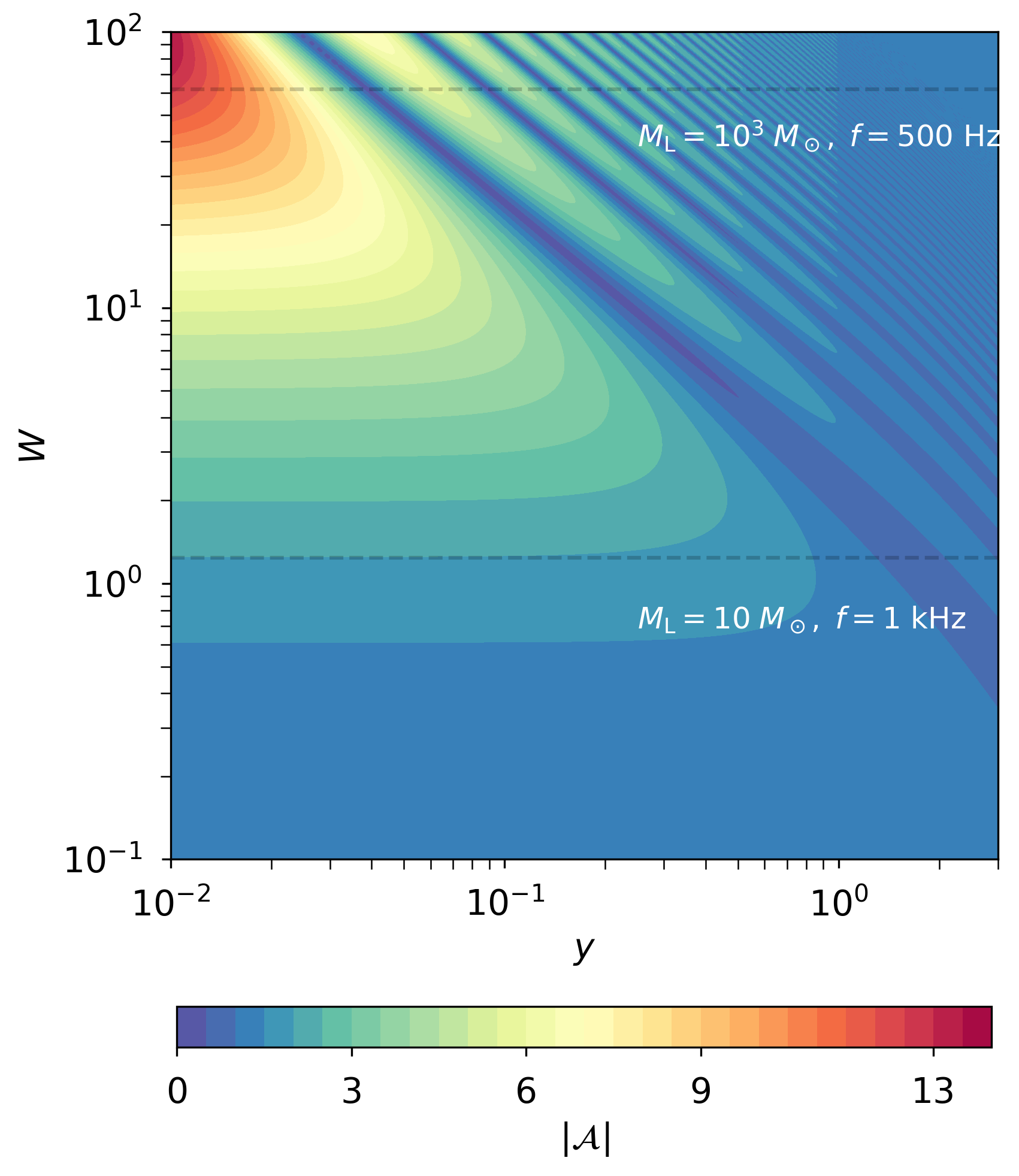}
\caption{The amplification factor ($|\mathcal{A}|$) is a function of $\w$ and varying $y$. At each value of $\w$, a degeneracy exists with multiple $\ml$ and $f$ combinations. The dashed line in the graph represents one of the possible combinations of $\ml$ and $f$ at that particular $\w$ value. When $\w < 1$, regardless of the value of $y$, the $|\mathcal{A}|$ remains small, indicating insignificant oscillatory behavior (Diffraction). For $\w > 1$, there is an increase in the $|\mathcal{A}|$ at a lower $y$ values, while a decrease is observed at higher $y$ values, indicating high oscillatory behavior (Interference).}
    \label{fig:amplification-pml-2d}
\end{figure}

When the wavelength of a signal is significantly smaller than the Schwarzschild radius of the lens, the condition is referred to as the geometric optics (GO) limit. Under this limit, multiple signals having a definite magnification arriving with respective time delays~\cite{10.1143/PTPS.133.137, Takahashi_2003, Petters}.  Conversely, when the signal's wavelength is comparable or larger than the Schwarzschild radius of the lens, one is in the wave optics (WO) limit, where a frequency dependent amplification of the waveform occurs.

% ================================== %
% Subsection
% ================================== %
\subsection{Lensing by a point mass in WO regime}
Our study focuses on the microlensing of continuous wave (CW) signals by objects with masses ranging from $1$ to $10^4$~$\,M_\odot$ consisting of  stellar objects including NS, BH, or IMBH,  in our Galaxy. These objects can be modeled as point mass lenses (PML) due to their concentrated mass distribution ~\cite{Gravitational_lenses1992,strong_weak_micro, 10.1143/PTPS.133.137}.

% ================================== %
% Figure: Amplification factor 1D 
% ================================== %
\begin{figure}[h]
    \centering
	\includegraphics[scale=0.5]{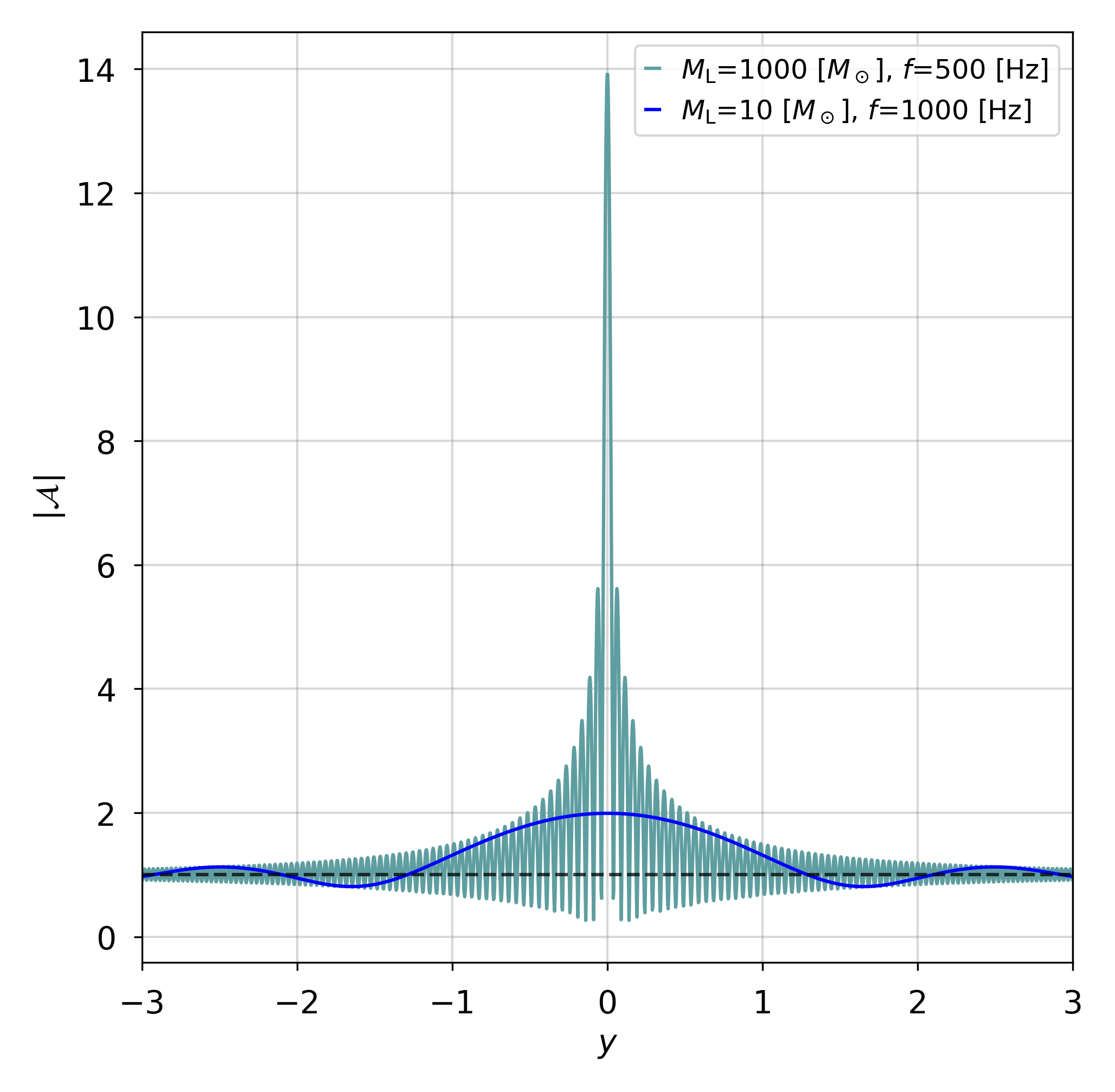}
    \caption{The amplification factor $|\mathcal{A}|$ varies with the parameter $y$, ranging up to [-3, 3] for two specific scenarios (i) $\ml =1000 \; M_\odot$, $f=500$ Hz, (ii) $\ml =10 \; M_\odot$, $f=1000$ Hz.  The first case have greater $\w=61.9$, leading to a higher $|\mathcal{A}|$ at lower $y$ and a more oscillatory nature compared to the second case $\w=1.2$. As $y$ increases, the value of $|\mathcal{A}|$ decreases in both cases. The dashed black line indicates the position where $|\mathcal{A}|=1$.}
    \label{fig:amplification-pml-1d}
\end{figure}

It is fortunate that PML is one of the lens models for which an analytical expression for the amplification factor can be derived. Solving the Fresnel-Kirchhoff diffraction integral \eqref{eq:Awy}  for the effective lensing potential $\psi(x) = \ln(x)$, the amplification factor for PML ~\cite{PhysRevLett.80.1138, 10.1143/PTPS.133.137, Takahashi_2003} is
\begin{eqnarray}
    \begin{aligned}
        \mathcal{A}(\w, y) &= \exp\left[ \frac{\pi \w}{4} + i\frac{\w}{2}
        \left( \ln \left(  \frac{\w}{2} \right)
        -2 \phi_m(y) \right)\right] \\
        &\times {_1}{F_1} 
        \left( \frac{i}{2} \w,1;\frac{i}{2} \w y^2 \right)
        \Gamma \left( 1- \frac{i}{2} \w \right), 
        \label{eq:pml-af}
    \end{aligned}
\end{eqnarray}
where ${_1}{F_1}$ is the confluent hyper-geometric function, $\Gamma$ is the Euler's gamma function. The constant $\phi_m(y)$, utilized to anchor the time-delay function reads: $\phi_m(y)=(x_m-y)^2/2-\ln{x_m}$, is the position of positive parity with a magnified signal in the context of GW lensing. In the case of PML model, the Einstein radius reads:
\begin{equation} 
    \re = \sqrt{\frac{4G \ml}{c^2} \frac{\dls \dol}{\dos}}
    \label{Einstein_radius}
\end{equation}
By taking the characteristic length scale (squared) as $\xi^2 _{0} = \re^2$, the dimensionless frequency becomes $\w = 8\pi G \ml f /c^3$, where $\ml$ is the mass of the lens. When $\w$ is significantly greater than one, i.e., higher lens masses and frequencies, Eq.~\ref{eq:pml-af} can be considerably simplified~\cite{PhysRevLett.80.1138, 10.1143/PTPS.133.137, Takahashi_2003} and falls into the GO regime. Given the mass range and frequencies in our analysis, we employed the full-wave optics expression in our work.

% ================================== %
% Figure: w, f and ML
% ================================== %
\begin{figure}[htp!]
    \centering
	\includegraphics[scale=0.5]{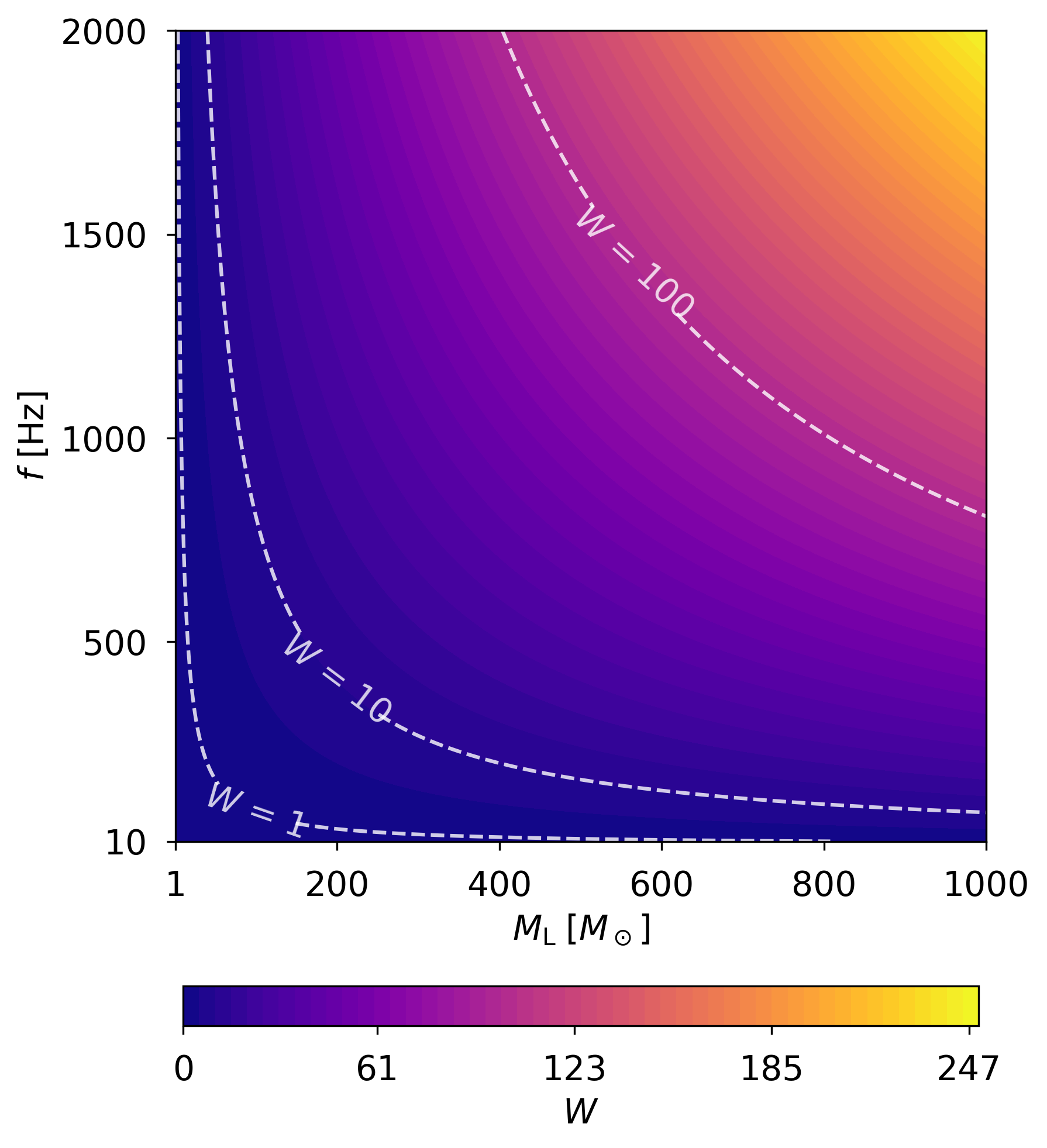}
    \caption{The correlation between the frequency ($f$) and the mass of lensing object ($\ml$) is explored for various values of $\w$. Since $\w \propto \ml$ and $f$, $\w$ exhibits degeneracy values for different combinations of $f$ and $\ml$. For example, the white dashed line in the graph represents the degeneracy value of $\w$ at specific points, namely $\w$ = 1, 10, and 100.}
    \label{fig:wfm_2d}
\end{figure}

The amplification resulting from a PML is a function of $\w$ and $y$, which is estimated using Eq.~\ref{eq:pml-af}. When the value of $\w \lesssim 1$, lensing  results in negligible amplification, irrespective of impact parameter $y$. This scenario is typical of the diffraction effect. On the other hand, when $\w \gtrsim 1$, $|\mathcal{A}|$ shows an oscillatory behavior, producing an interference pattern. Furthermore, when the value of $y$ is small, $|\mathcal{A}|$ is characterized by high magnification; however, this magnification diminishes as $y$ increases. There is a transition region between $\w \gtrsim 1$ and $\w \ll 1$, where the amplitude and oscillation vary gradually. The two cases of $\w$ that involve $y$ and $|\mathcal{A}|$ are shown in Fig.~\ref{fig:amplification-pml-2d}. The diffraction and interference patterns hereafter are referred to as \textit{lensing patterns}.

The $\w$ depends on $\ml$ and $f$, and different combinations of the $\ml$ and $f$ values can result with the same $\w$ and $\mathcal{A}$ at a given $y$ leading to degeneracy between these variables. A lensing pattern for two specific examples of $\w$; (i) $\ml =1000 \; M_\odot$, $f=500$ Hz, $\w=61.9$ (ii) $\ml =10 \; M_\odot$, $f=1000$ Hz, $\w=1.2$, both of which represented with three-dimensional information in Fig.~\ref{fig:amplification-pml-2d}, that is unfolded into the two-dimensional information $|\mathcal{A}|$ versus $y$, as shown in Fig.~\ref{fig:amplification-pml-1d}. The first example shows a larger $\w$ than the second example, leading to a greater $|\mathcal{A}|$ value at a lower $y$. Additionally, the larger $\w$ displays a more oscillatory nature than the smaller $\w$. As $y$ increases, the value of $|\mathcal{A}|$ decreases in both examples. It is worth noting that within this oscillatory pattern, $|\mathcal{A}|$ can fall below one, indicating that the signal might be de-magnified for certain values of $y$.

Further in the paper, we explore the parameter space for $\ml$, $f$, and $\w$ to identify the significant amplification factor $|\mathcal{A}|$ associated with $y$. The frequency range is set to [$10$, $2000$]~Hz, which is the predominant sensitivity range for CW signal in the second-generation detector~\cite{2024arXiv240902831S}. The mass range for the analysis is limited to [$1$, $10^4$]~ $M_\odot$. To illustrate the impact of $\ml$ on $\w$, the maximum value for $\ml$ is set at $10^3$ as shown in Fig.~\ref{fig:wfm_2d}. Since $\w$ is directly proportional to the $\ml$, an increase in lens mass leads to an increase in $\w$, at given $f$.

% ================================== %
% Subsection %
% ================================== %
\subsection{Basics of microlensing}
Strong lensing occurs when the alignment between the source and the lens is nearly perfect along the line of sight. In physical terms, this implies that the projected position of the source must fall within the Einstein radius of the lens. Microlensing is a specific instance of strong lensing that occurs when the Einstein radius of the lens is relatively small (of the order of milli/micro-arcseconds) for galactic lensing objects. It is most unlikely to observe the multiple signals (such as the two signals seen in the PML-GO case) separately;  instead, they merge into a single signal, causing the overall observed signal to be magnified. Consequently, the strength of the alignment implies that the relative transverse motion of the source concerning the lens could manifest itself. Namely, the $y$ parameter will change in time as the source moves closer to the line of sight, amplifying its signal strength until it reaches its closest point of encounter at $y_0$. As the source moves away, the signal will gradually return to its baseline strength.

The relative motions of the lens, the observer, and the source (projected on the sky) can be approximated as linear when then the light curve is symmetric. This is the essence of microlensing in the EM domain envisaged by Paczy{\'n}ski~\cite{Paczynski} and routinely observed in EM~\cite{Mroz2024}. We aim to study an analog of this well-known phenomenon within the GW domain, going beyond the GO formalism framework traditionally applied in EM microlensing. Instead, we broaden our analysis to include the WO regime, which is more appropriate for realistic GW microlensing scenarios.

Let us emphasize subtle differences between the gravitational lensing of short-transient (e.g., CBCs) and CW signals. First, the influence of impact parameters $y$ due to relative transverse motions. In the short-transient signals in the LVK frequency band, $y$ changes insignificantly and remains constant. On the other hand, the CW signal $y$ becomes noticeable during the observation period and varies over time. The second difference is that frequency changes rapidly in a short-duration transient signal (chirp). Accordingly, for a given $M_L$, the $\w$ parameter also varies. In some cases, one can experience the transition between GO and WO regimes, a phenomenon that remains to be studied. In contrast, CW signals are monochromatic. Therefore, for a given $M_L$, the $\w$ parameter is fixed, and the amplification factor can change only due to the time-varying impact parameter $y(t)$.

% ================================== %
% Figure: Lens plane
% ================================== %
\begin{figure}[htp!]
    \centering
    \includegraphics[trim={0 20pt 0 20pt},clip,width=0.8\columnwidth]{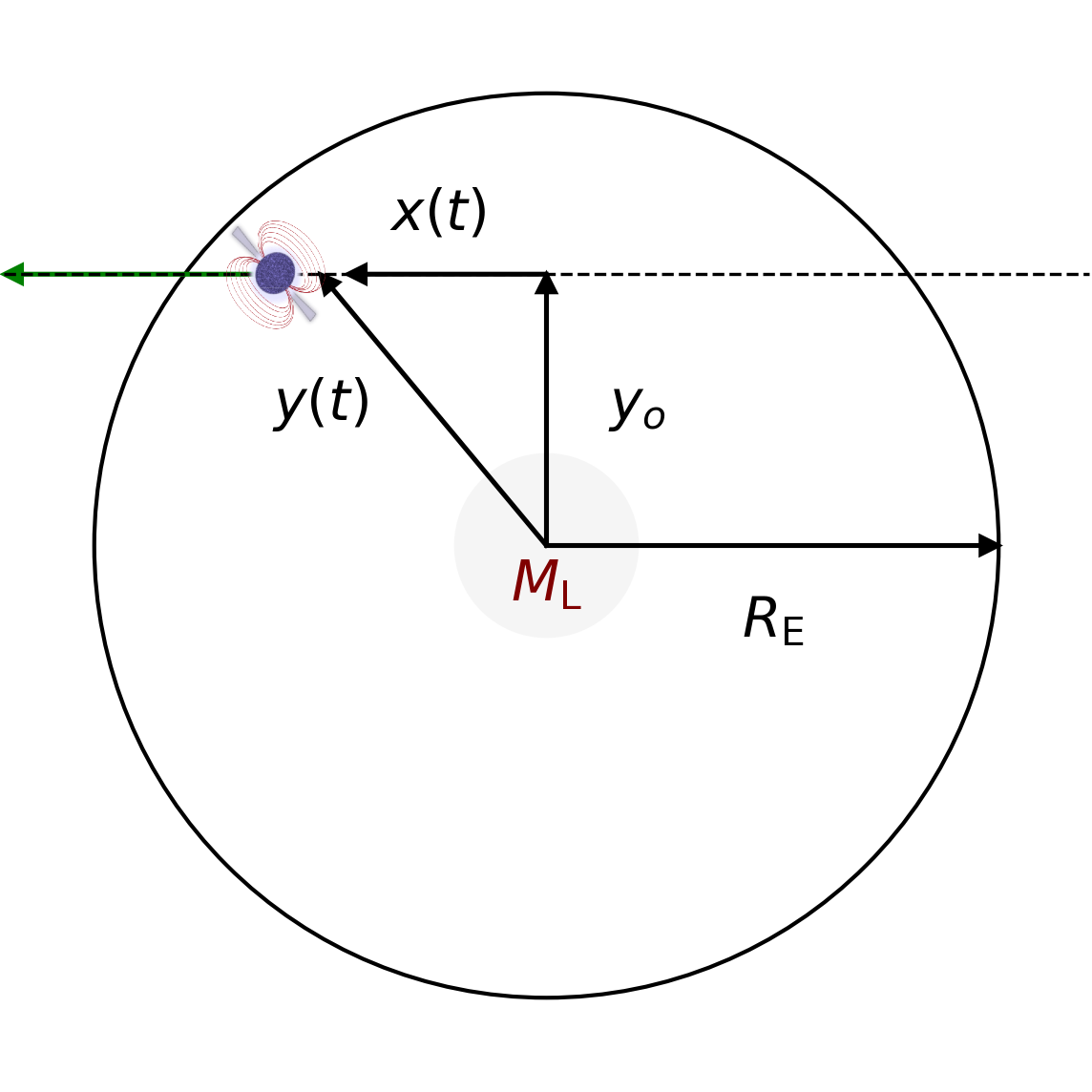}
    \caption{Schematic scenario for a stationary lens and a source moving transversely $x(t)$ from right to left in the line of sight, crossing the Einstein radius $\re$. The closest distance between the lensing object and the source at time $t_0$ is denoted by $y_0$.}
    \label{fig:illust_lens-plane}
\end{figure}

To simplify the scenario, we can choose a reference frame defined by the line of sight to the lens and the lens plane, which is orthogonal to it. We assume the source moves with a transverse relative velocity and the lensing object is stationary, as illustrated in Fig.~\ref{fig:illust_lens-plane}. The impact parameter $y(t)$ at some moment of time~\cite{Liao_2019, intro_to_lensing} can be written as:
\begin{eqnarray}
    y(t) = \sqrt{y_0^2 + x^2(t)},
    \label{micorlensing_y(t)}    
\end{eqnarray}
where $x(t) = (t-t_0)/{\te}$, $y_0$ is the closest distance between the lens and the source at time $t_0$. The $\te = R_E/\vrel$ is the Einstein crossing time and is the time taken by the source to cross the Einstein radius $\re$ of the lens. $\te$ is influenced by the distances $\dol$, $\dos$, the relative velocity $\vrel$, and the mass of the lensing object in the lensing system. We define $\dr = \dol / \dos$ as the ratio of respective distances and $\dos > \dol$. It is convenient to explore lensing parameters by fixing the distance $\dol$ or $\dos$. For example, if $\dr$ is 0.5 at a given $\dol$, $\dos$ is twice the $\dol$. Remembering the formula Eq.\ref{Einstein_radius} for the Einstein radius $\re$ of PML one can express $\te$ in terms of $\dr$ and $\dol$
\begin{eqnarray} 
    \te &=& \frac{1}{\vrel} \sqrt{\frac{4 G \ml}{c^2} \dol (1 - \dr)}.
    \label{eq:einstein-crosstime}
\end{eqnarray}

The above equation can be expressed in terms of $\dos$ also, by substituting $\dol = \dr \dos$. $\te$ is an essential quantity allowing us to assess the feasibility of detecting the lensing patterns during the observation time $\tobs$. Since the lensing system we are studying is located within the galaxy, we can constrain the parameters that rely on estimating the $\te$. We have a good measurement of the size and kinematics of our Milky Way galaxy. Its diameter is $\approx28$~kpc~\cite{Goodwin-1998O}, and the distance from the Solar System to the galactic center is about $8.5$~kpc~\cite{Gillessen_2017}. So, we pick up the parameters for the realistic microlensing GW scenarios: $\dol \in$ [0.01, 11]~kpc, and $\dr \in [0.5, 1)$.  

% ================================== %    
% Figure: Einstein Crossing Time
% ================================== % 
\begin{figure}[htp!]
    \centering
	\includegraphics[scale=.5]{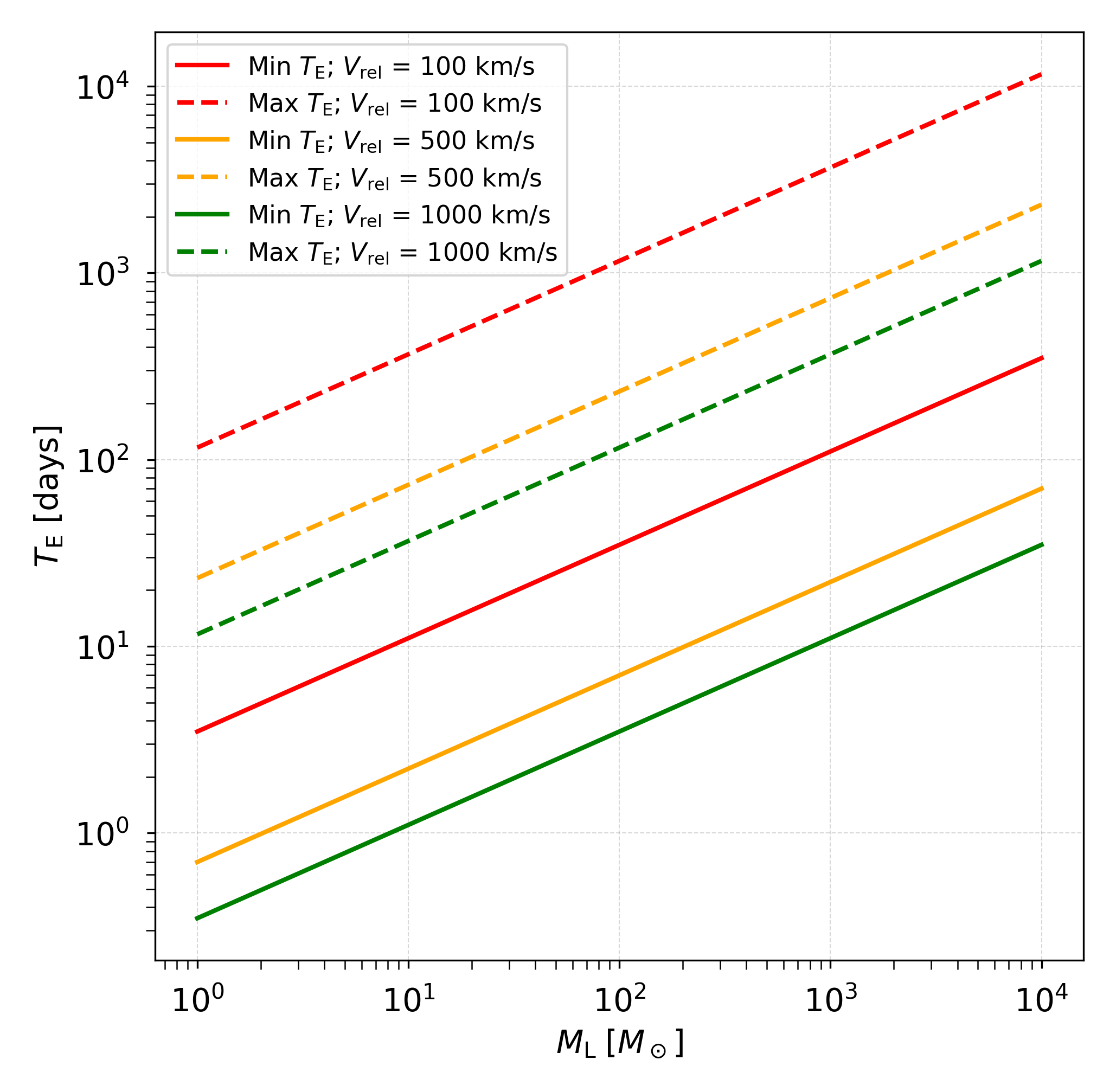}
    \caption{Estimates for the minimum and maximum Einstein crossing times have been calculated across a range of lens masses. These estimates account for the distance from the observer to the lens, which spans from 10~pc to 11~kpc, and the distance from the observer to the source, ranging from 10~pc to 22~kpc. The parameter $\dr$ is also set [0.5, 1), considering three different velocities. At a specific distances and $\vrel$, $\te \propto \sqrt{\ml}$. The dashed line represents minimum $\te$, and the solid line represents maximum $\te$.}
    \label{fig:te}
\end{figure}

The rotation speed of the Milky Way's disk stars averages around $V_{\mathrm{rot}} = 220$~km/s, with a root mean square (RMS) deviation of $V_{\mathrm{RMS}} = 50$~km/s relative to this speed \cite{Galactic_dynamics}. In contrast, the motion of stars in the galactic bulge is randomly oriented, lacking any net rotation, and displays a velocity dispersion of $V_{\mathrm{RMS}} = 150$ km/s \cite{Galactic_dynamics}. These factors have the following consequences regarding possible values of relative transversal velocities in lensing systems. If both the source and the lens are positioned within the disk on the same side relative to the galactic center, the $\vrel$ can vary between $50$~km/s and $200$~ km/s. For sources situated in the bulge, a higher $\vrel$ of 200-300 km/s is typical. Conversely, if the source is located on the opposite side of the lens with respect to the galactic center, the relative velocity can reach values as high as $\vrel \sim$ 400-500 km/s. From the perspective of an all-sky, all-frequency lensing search, we take into account all plausible lensing scenarios and set the $\vrel$ range from $100$~km/s to $1000$~km/s. The upper limit of this range accounts for potential instances where the NS source possess that kick velocity~\cite{Chatterjee_2005, Chatterjee_2004, Bruzewski_2023, verbunt, sartore, 10.1093/mnras/staa958, Lai2001}. 

We estimate the minimum and maximum Einstein crossing times at three relative velocities $\vrel$ = 100, 500, 1000 km/s in the $\dol$, $\dos$, and $\dr$ parameter space considered in our studies, which are mentioned earlier and shown in Fig~\ref{fig:te}. $\te$ can also be written in physical units to help build intuition about the feasible time scale of lensing~\cite{Liao_2019}. $\te \propto \sqrt{\ml}$ can be scaled for different $\ml$ at given distances, and $\vrel$; $\te$ increases as the $\ml$ increases. This means that a larger $\ml$ results in a larger $\re$ and increases the time duration for the source to cross the $\re$. Moreover, the $\te$ diminishes as the source's velocity increases for a fixed $\ml$ and distances. Studying the $\te$ with associated parameters gives an insight into determining the observational possibility of the lensing effect in second-generation detectors. 

The lensing system parameters associated with the lensing probability for a galactic lensing object and CW sources have not been extensively studied yet. However, the lensing probability for the supermassive black hole at the center of our galaxy acting as a lens for the neutron stars located behind it was studied ~\cite{Paolis, Basak_2023}. Moreover, the different lens models, such as the Singular Isothermal Sphere (SIS) model ~\cite{Liao_2019}, yield differing lensing probabilities when applied to globular and open clusters in our galaxy that act as a lens. It is essential to note that the estimated lensing probability varies depending on the model employed, whether it is the model for the population of sources ~\cite{Marek, 10.1093/mnras/staa073} or the lens.

Regardless of the unexplored lensing probability, typical promising scenarios for investigating this phenomenon include galactic centers and globular clusters. We explored the lensing parameter space more generally for all galactic CW sources and lenses in the case of $\tdfstat$ analysis, as a proof of concept, we select specific parameters that correspond to the sensitivity of second generation detectors and estimate the amplification factor using Eq.~\ref{eq:pml-af}. The following section discusses the prospects for detecting lensed CW signals in a noisy data and comparing them with unlensed signals. We inject a signal with a specific amplification level into Gaussian noise for a range of parameters and recover it using the {\tdfstat} method, which employs a matched filter technique.
% ================================== %    
% TDFstat
% ================================== % 
\section{\label{sec:tdfstat} Signal analysis with the Time-Domain F-statistic method}
\subsection{TDFstat method}
Asymmetric rotating NSs, ultralight scalar bosons surrounding spinning BHs, and primordial BH binaries with planetary or asteroid masses in their inspiral phase are potential sources of CW signals. These are detectable sources for second-generation detectors~\cite{Riles2023, Haskell2023, WETTE2023102880}. In order to introduce a search method suitable for CWs, one needs to briefly describe the CW signal characteristics. Without loss of generality, let us assume the source to be an asymmetric NS, rotating about one of its principal axes of inertia. In the detector frame, the GW amplitude~\cite{PhysRevD.58.063001} is given by
\begin{eqnarray}
    \begin{aligned}
        h(t) &=h_0 \Bigl(F_+(t,\alpha,\delta,\psi)\frac{1+\cos^2{\iota}}{2} \cos{\phi(t)} \Bigr.\\
    &+ \Bigl. F_{\times}(t,\alpha,\delta,\psi) \cos{\iota} \sin{\phi(t)} \Bigr),
    \label{eq:hot}
    \end{aligned}
\end{eqnarray}
where $h_0$ is the intrinsic amplitude of the signal, $F_+$ and $F_{\times}$ are the antenna patterns of the detector, $\delta$ and $\alpha$ denote sky position of the source (declination and right ascension), $\psi$ is the polarization angle, $\iota$ is the angle between the total angular momentum vector of the NS and the direction from it to the Earth, and $\phi(t)$ is the phase of the signal. For an almost monochromatic CW signal around a frequency $f$, evolving with a frequency time derivative $\dot{f}$, the phase is well approximated by a Taylor expansion, 
\begin{eqnarray}
    \begin{aligned}
        \phi(t) = \phi_0 + 2\pi f \Bigl( t + \frac{{\bf n} \cdot {\bf r}_d(t)}{c} \Bigr) \\ + 2\pi \dot{f} \left (t + \frac{{\bf n} \cdot {\bf r}_d(t)}{c} \right)^2,
    \label{eq:phaseevo}    
    \end{aligned} 
\end{eqnarray}
where $\phi_o$ is an initial phase. In order to take into account the movement of the detector with respect to the source, one usually evaluates the phase in the Solar System Barycenter (SSB); thus, we define the vector ${\bf r}_d(t)$ joining the SSB with the detector, and the unit vector ${\bf n}$ pointing from SSB to the source. The amplitude $h_0$ is given by 
\begin{eqnarray}
    \begin{aligned}
        h_0 &= \frac{4\pi^2G}{c^4} \frac{\epsilon I_{zz} f^2}{d}, \\
        & \approx 1.06\times10^{-26} \left(\frac{\epsilon}{10^{-6}}\right) \\
        & \times\left(\frac{I_{zz}}{10^{38}\>\rm{kg\ m}^2}\right)\left(\frac{f}{100\>{\rm  Hz}}\right)^2
\left(\frac{1\>{\rm kpc}}{d}\right), 
    \label{eqn:hexpected}
    \end{aligned} 
\end{eqnarray}
where $d$ is the distance from the detector to the source, $\epsilon=(I_{xx}-I_{yy})/I_{zz}$ is the deviation from the symmetry of the NS, sometimes called the ellipticity. $I_{xx}$, $I_{yy}$ and $I_{zz}$ are the principal moments of inertia. 

%-------------------------%
% Figure: TDF Illustration
%-------------------------%
\begin{figure}[htp!]
	\includegraphics[scale=.5]{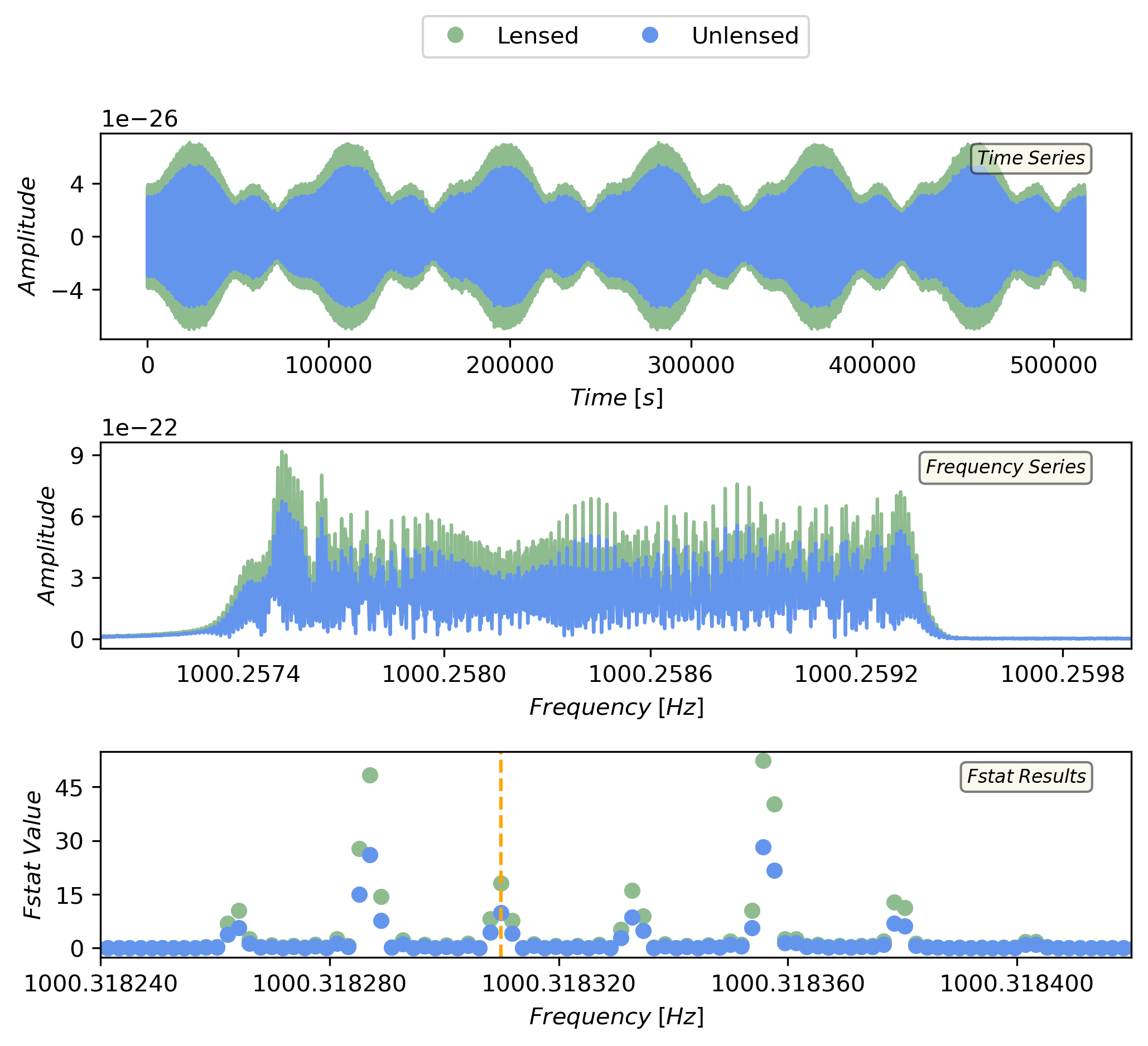}
    \caption{An example of a simulated CW signal, with the unlensed signal in blue and the lensed signal in green. \textit{Top panel:} six-day time-series simulation with Gaussian noise of mean $\mu =0$, standard deviation $\sigma_{\mathrm{n}} = 0.0002$, the noise strain amplitude is $n_{\mathrm{o}}=10^{-23}$, and the signal strain amplitude is $h_{\mathrm{o}}=10^{-25}$. \textit{Middle panel:} The frequency domain representation of the above highlights significant amplitude, phase, and frequency shifts induced by modulations, such as Doppler effects, sidereal variations, and the detector's response, evident in both the unlensed and lensed signals. \textit{Bottom panel:} Distributions of {\fstat} values, evaluated using the {\tdfstat} analysis, corresponding to the two injected signals. The distance between the {\fstat} peaks equals one Earth sidereal day frequency; heights of the peaks contain information of each signal's parameters. The orange dashed vertical line is the injected signal frequency at $f=1000.31830988$~Hz with $\dot{f}=0$ (see Appendix~\ref{app:App-B} for more details of the signal parameters). The {\fstat} values are higher for the lensed signal, as expected for the microlensing amplification. In subsequent Figs.~\ref{fig:snr-segment1} and \ref{fig:snr-segment6}, we select the maximum value of the {\fstat} distribution, for each time-domain segment results.}
    \label{fig:td_fd_illustration}
\end{figure}
For very weak signals hidden in the noise the standard technique of detection is the use of matched filter. In the {\tdfstat} method the `quality' of match between the template (CW signal with all modulation for given parameters) and detector's data is evaluated. The $\mathcal{F}$-statistic \cite{PhysRevD.58.063001,Astone:2010} is obtained by analytically maximizing the likelihood function with respect to the four unknown signal parameters, characterizing the template - the polarization $\psi$, angle $\iota$, amplitude $h_0$ and the phase of the signal $\phi_o$, while the other four parameters - the sky position $\delta$ and $\alpha$, and frequency parameters, $f$ and $\dot{f}$ - form the actual search parameter space (see Sec. IIIA in~\cite{PhysRevD.58.063001}). The {\fstat} value is directly related to the SNR $\rho = \sqrt{2\mathcal{F}-4}$. The SNR scales with both the data duration also called segment duration time $\ts$ and the amplitude of CW signal $h_0$, $\rho = h_0\sqrt{\ts/ \tilde S_f}$, where $\tilde S_f$ is the power spectral density of the detector's noise at frequency $f$. The amplitude $h_0$ is thus directly linked with the value of $\mathcal{F}$ (for more details, see \cite{PhysRevD.58.063001}).

A typical application of the {\tdfstat} pipeline is an all-sky, broad frequency range search for CW signals with {\em a priori} unknown parameters. The first stage is a coherent {\fstat} search on a long duration ($\ts$ order of few days) time-domain data, centered around a given frequency $f$, on a grid of $(\delta, \alpha, f, \dot{f})$ values. For computational feasibility the frequency range is divided in narrow frequency bands, with typical bandwidth of 0.25 - 1 Hz. The {\fstat} values and candidate signal parameters obtained in a coherent search across a set of time-domain segments comprising the observing run are then used during a coincidences search between the signal candidates from different time segments. This procedure is used to select significant signal candidates according to specific criteria, e.g., persistence throughout a sequence of time segments or, as in the case considered here, a lensing pattern of $h(t)$ in a sequence of time segments, which may be caused by temporal amplification of $h(t)$ due to the lensing.
% ================================== %    
% Subsection
% ================================== % 
\subsection{\label{sec:data-preparation} Data preparation}
The two separate time-domain datasets of simulated signals have been created: unlensed and lensed CW from NS. Both sets of signals included same Gaussian noise $n(t)$. The unlensed continuous wave signal $h(t)$ with noise is represented as
\begin{eqnarray}
    s(t) = h(t) + n(t)
    \label{eq:hot_unlensed}
\end{eqnarray}
Similarly, the lensed CW signal $h^l(t)_{\mathrm{lensed}}$ with noise is expressed as\footnote{The superscript '$l$' indicates a lensed signal; if there is no superscript, it refers to an unlensed signal.}
\begin{eqnarray}
    s^l(t) = h^l(t) + n(t).
    \label{eq:hot_lensed}
\end{eqnarray}
%-----------------------------%
% Figure: Analytical Expression
%-----------------------------%
\begin{figure*}[htp!]
    \centering
	\includegraphics[scale=.37]{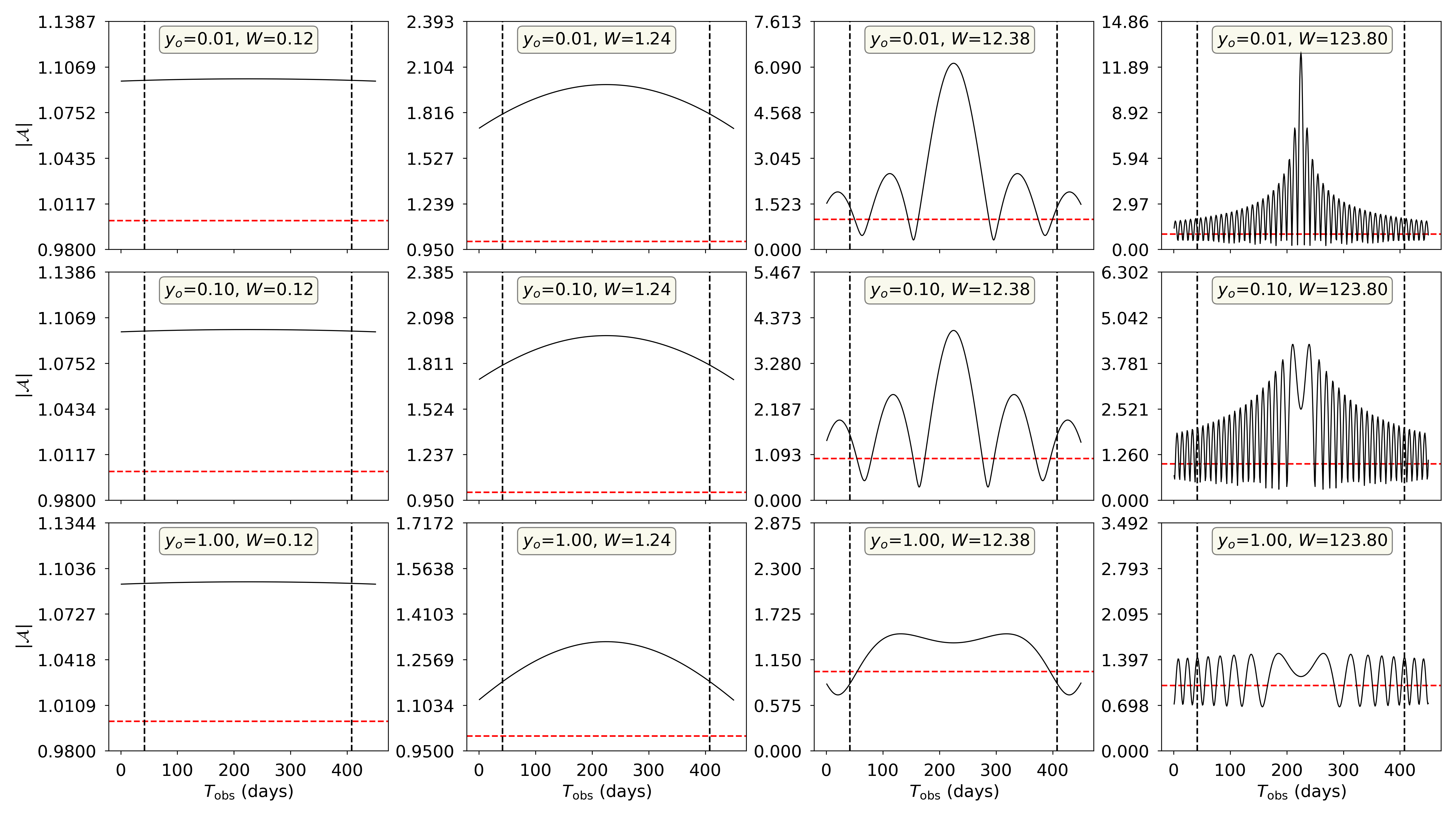}
    \caption{An analytical estimation of the amplification factor and corresponding lensing patterns for twelve parameter sets is based on combinations of the three closest impact parameters, $y$ (0.01, 0.1, and 1), along with four $\w$ parameters (0.12, 1.24, 12.38, and 123.80). The horizontal red dashed line indicates an amplification factor of one. The region between the vertical black dashed lines represents one year of observation time. When $\w=0.12$ (\textit{Column 1}), the amplification shows minimal variation across different $y$ values, resulting in a nearly flat response without any noticeable lensing pattern. In the case of $\w=1.2$ (\textit{Column 2}), a lensing pattern becomes apparent, exhibiting slight increases and decreases in amplification. For the $\w=12.38$ (\textit{Column 3}) and $\w=123.80$ (\textit{Column 4}), amplification significantly increases compared to earlier values of $\w$; however, amplification decreases as $y$  increases, and this trend is consistent across other values of $\w$ as well. Notably, lensing patterns are significant in both cases. It is important to mention that the fringe patterns at $\w$ = 123.80 are distinct from those at $\w$ = 12.38.}
    \label{fig:af-te}
\end{figure*}

%----------------------------------------%
% Figure: TDF one day time series segment
%----------------------------------------%
\begin{figure*}[htp!]
	\includegraphics[scale=.45]{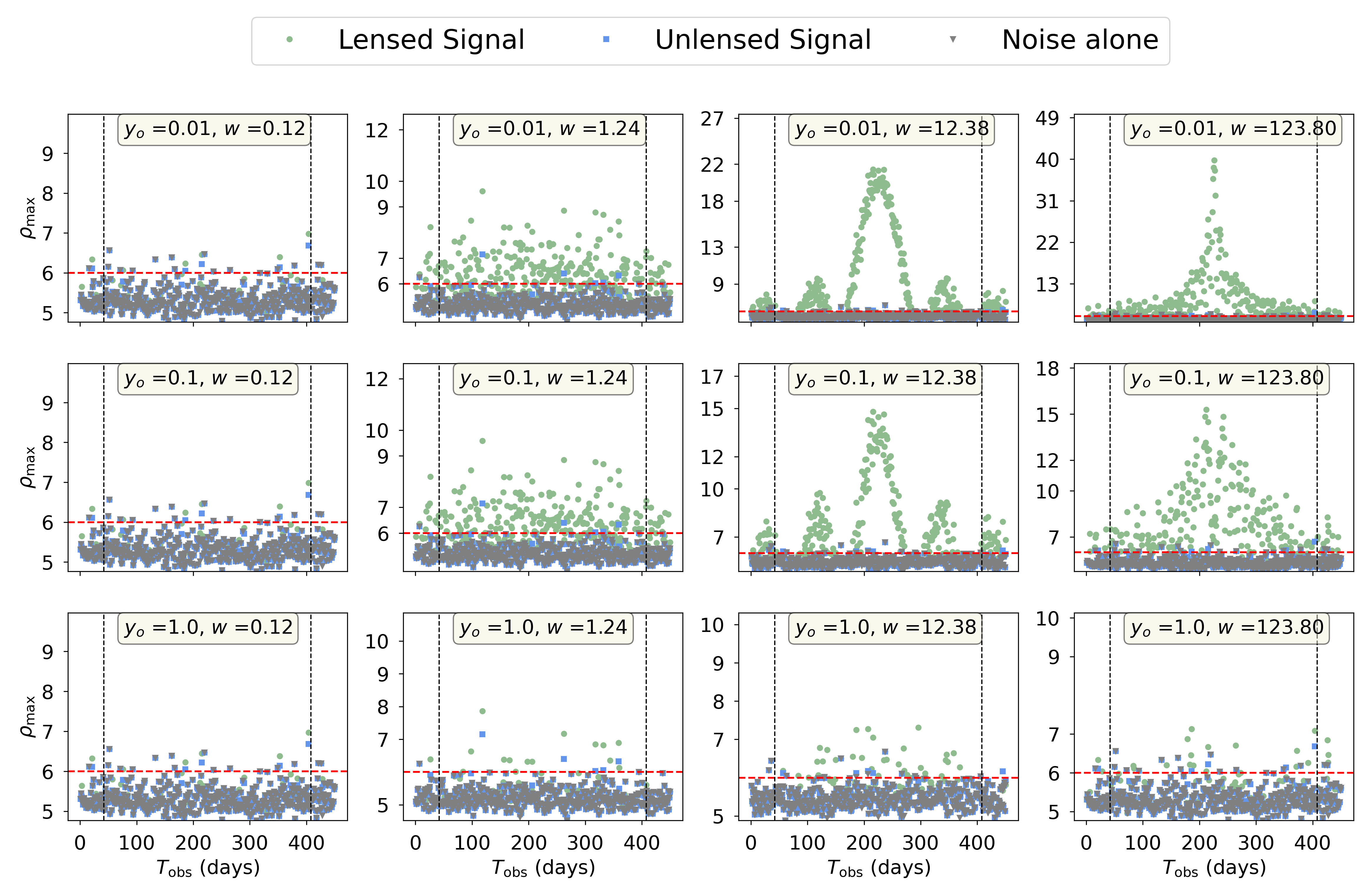}
    \caption{The SNR was estimated from the  {\tdfstat} analysis for each one-day time series segment over 450 days. The SNR of lensed signals is represented by green dots, unlensed signals by blue squares, and noise-only signals by gray triangles. The dashed red line indicates the SNR threshold, which is set as a reference. Most SNR values for noise-only and unlensed signals are below the SNR threshold of 6. For lensed signals, the impact parameter $y_o=1$ (\textit{Bottom row}), most of the SNR also remains below the reference threshold of SNR 6, regardless of the value of $\w$. Nevertheless, for $\w=1.24$ and $y_o=0.01$ and $.1$, the SNR for the lensed signal exceeds the SNR threshold. The SNR exhibits a lensing pattern at $\w=1.24$ and $\w = 123.80$ for $y_o = 0.01$ and $y_o= 0.1$, displaying distinct features. The area between the vertical black dashed lines represents one year of observation time. Refer to Appendix~\ref{app:App-B} for details about parameters.
    }
    \label{fig:snr-segment1}
\end{figure*}
We assumed that the source's frequency is constant and has no other frequency modulations prior to lensing; spin-down effects are ignored and set zero in the simulation. Since, in general, the position of the source will change over time due to relative movements, it will result in a time-dependent impact parameter $y(t)$. Thus, the amplification factor for the Point Mass Lens (PML) in Eq.~\ref{eq:pml-af} can be expressed as a time-dependent amplification factor at given $f$, denoted as $\mathcal{A}(f, t)$. As a result, the lensed signal is the product of the gravitational lensing amplification factor and the CW signal   $h^l(t) = A(f, t) \; h(t)$; see Appendix~\ref{app:App-A} for details.

%----------------------------------------%
% Figure: TDF six day time series segment
%----------------------------------------%
\begin{figure*}[htp!]
    %\raggedleft
	\includegraphics[scale=.45]{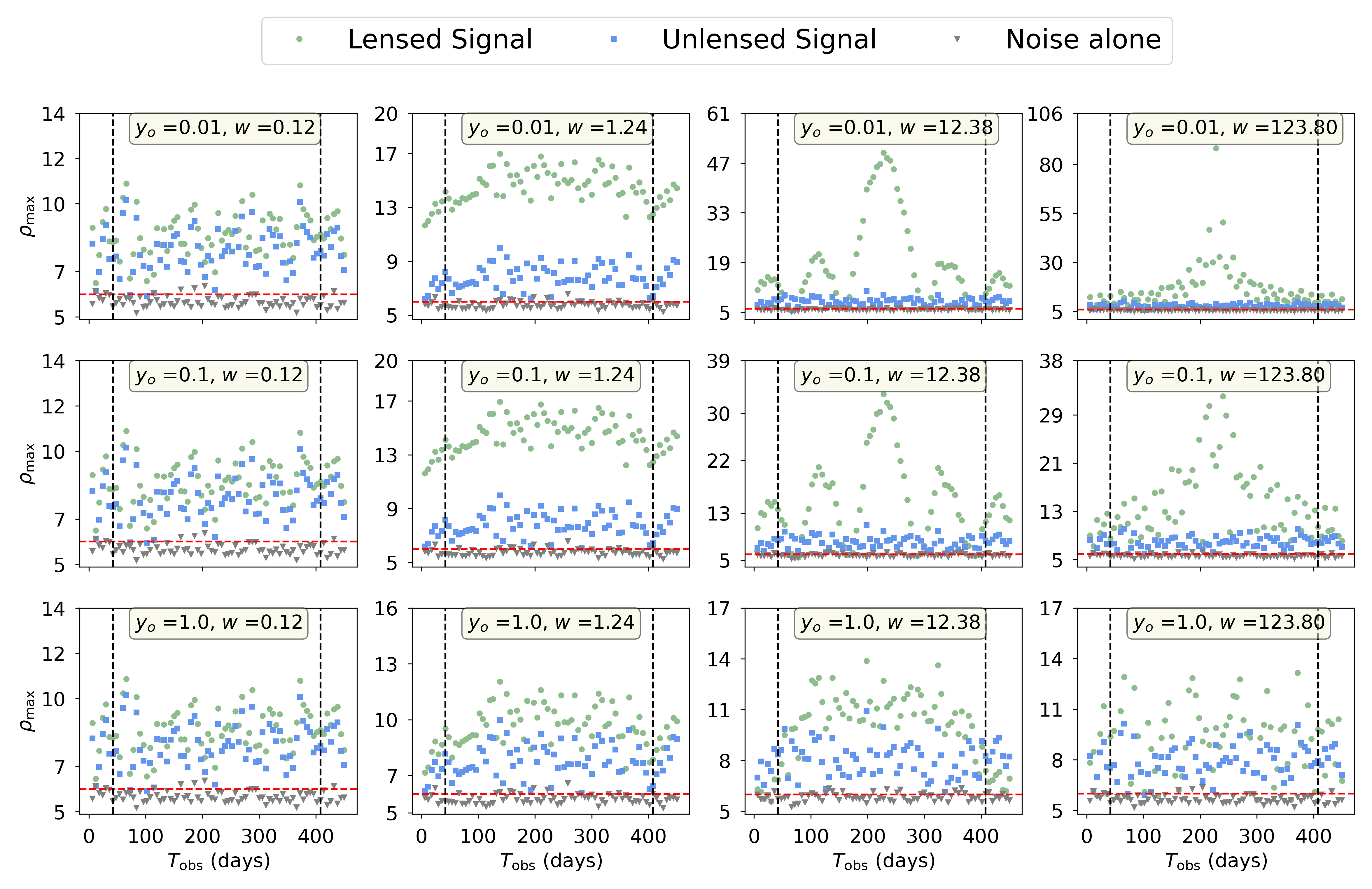}
    \caption{The SNR was estimated from the {\tdfstat} analysis for each six-day time series segment over 450 days. Lensed signals are marked by green dots, unlensed signals by blue squares, and noise-only signals by gray triangles. The dashed red line indicates the SNR threshold, which serves as a reference point. Most SNR values for noise-only signals are below the SNR threshold of 6. The SNR for both unlensed and lensed signals exceeds the established threshold, and the SNR for the unlensed signal remains nearly constant across all cases. At $\w=0.12$ (column 1), the SNR for the lensed signal is slightly higher than that of the unlensed signal. At $\w=1.24$ (column 2), there is a noticeable difference between the two. When $\w=12.38$ (column 3) and $\w=123.80$ (column 4), the SNR exhibits a lensing pattern; however, this pattern loosely appears due to the limited number of data points in the six-day semi-coherent analysis. Additionally, both the unlensed and lensed SNRs decrease as $y_o$ increases. The area between the vertical black dashed lines represents one year of observation time. Refer to Appendix~\ref{app:App-B} for details about parameters.}
    \label{fig:snr-segment6}
\end{figure*}
We have taken into account the Doppler effect caused by Earth's orbital motion around the Sun, as well as the modulation effects of Earth's rotation and antenna response for both sets of signals. In addition, we included noise-only data without any modulation for our analysis. By applying the Solar System Barycenter (SSB) reference frame, we can correct for frequency changes, ensuring that our analysis accurately reflects the measured signals. This correction is essential when employing the {\tdfstat} analysis~\cite{PhysRevD.58.063001, PhysRevD.59.063003}. The analysis is performed on a grid that serves as an efficient template bank~\cite{Pisarski_2015}, particularly for the Hanford detector.

After analyzing the lensing parameter space in Section~\ref{sec:lensing_intro}, we selected a specific subset of lensing parameters, which includes three frequencies and three lens masses for simulation, see Appendix~\ref{app:App-B} for details. The reference strain for the simulated CW signals is set to $h_{\mathrm{o}}=10^{-25}$, based on the upper limits of $h_{\mathrm{o}}$ set in the 100-200 Hz range during the first three LVK observing runs~\cite{Abbott_202208}. In the following, we adopt for the standard deviation of the noise the value  $\sigma_{\mathrm{n}} = 0.0002$, the representative noise strain amplitude $n_{\mathrm{o}}=10^{-20}$, based on the current overall quality of the data. Although the frequencies we considered fall within three different narrow bands of the sensitivity curve, we study the lensing effect using the same noise standard deviation and strain across all bands in our simulations. Altering the noise strains amplitude or standard deviation values affects the SNR and the lensing pattern.

The time duration to cross the Einstein ring is set to 366 days, which is close to the duration of the O3 observation run. Additionally, our simulation incorporates an extra 42 days for the time period, both prior to and following the crossing time. This extended time frame allows us to study the anticipated lensing effect, such as the increase of amplification as the source approaches the Einstein Radius and the decrease of amplification as the source moves away from it. Combining $\te$ with a relative velocity, $V_{\mathrm {rel}}$, we can estimate the effective distances $D$ for each lens mass, as mentioned in Eqn~\ref{eq:einstein-crosstime} and refer Appendix~\ref{app:App-B} for $\dr$ and $\vrel$ values. Three impact parameter, $y_\mathrm{o}$, values of 0.01, 0.1, and 1 are used to determine the amplification factor of the lensing signal. Figure~\ref{fig:td_fd_illustration} illustrates an example of a six-day time-series simulation with Gaussian noise, along with its Fourier transforms and the output from TDFstat.

% ================================== %    
% Subsection
% ================================== % 
\subsection{\label{sec:tdfstat-results} {\tdfstat} results}
Prior to conducting the {\tdfstat} analysis on the prepared data, we estimated the analytical amplification factor and analyzed the lensing pattern within the same parameter space as stated in the previous sub-section. Figure~\ref{fig:af-te} shows an analytical prediction of the lensing patterns that emerge from the microlensing effect for 12 parameter sets. With the increase of the closest impact parameter, $y_o$, relative to the position of the lensing object, the amplitude of the lensing pattern decreases. Additionally, the amplification factor is affected by the parameter $\w$. In Fig.~\ref{fig:af-te}, column 1 shows that when $\w \lessapprox 1$, the amplification slightly increases with the closest impact parameter $y_o$. When $\w$ approaches 1, the amplification factor exhibits a noticeable rise and fall, as shown in Fig.~\ref{fig:af-te}, column 2. Further increase of $\w \gtrapprox 10$ makes the interference patterns evident and distinguishable for different impact parameter values, each exhibiting a characteristic interference pattern. We aimed to consistently compare $\mathcal{A}$ across various $\w$ and $y$ with fixed $\tobs = 450$~days. The area between the vertical dashed black lines represents a duration of 366 days, while the horizontal dashed red line indicates a reference value $\mathcal{A}=1$. The observed lensing pattern from the microlensing effect is based on two cases involving $\te$ and $\tobs$.

$\te \leq \tobs$: In this case, the symmetrical lensing pattern (Fig.~\ref{fig:af-te}) resulting from the microlensing effect can be observed. This is similar to the Paczy{\'n}ski curve in EM lensing observations \cite{Paczynski}.

$\te > \tobs$: only a portion of the microlensing effect is visible. The observation would capture only partially the magnification of the source signal, and whether the amplification increases or decreases indicates the direction of the source's movement within the lens plane.

A set of lensing parameters is used to generate the lensed and unlensed versions of the signals, and the same set is used to evaluate the analytical amplification factor. Subsequently, the maximum SNR is calculated from the {\tdfstat} analysis for one-day and six-day time series segments over 450 days for each simulated dataset, i.e., unlensed, lensed, and noise-only are shown in Figs.~\ref{fig:snr-segment1} and~\ref{fig:snr-segment6}.
We set SNR threshold to 6, which is indicated by the dashed red line. In our simulated data, the noise alone contributes to SNR $\approx 6$. The unlensed signal, which has a signal amplitude of $h_{o}=10^{-25}$, also has a SNR $\approx 6$ in the one-day time series segment analysis. This outcome agrees with our simulation expectations due to the noise variance considered in the narrow-band analysis. Since the detector's sensitivity varies across frequencies, applying the same noise sensitivity to the considered frequencies can introduce a bias. Nevertheless, using the reference signal amplitude of $10^{-25}$ is relevant, given the upper limit of $10^{-25}$ at the 100 Hz frequency in the second-generation detectors.

Figure~\ref{fig:snr-segment1} shows that the amplitude of the lensed signal increases as the variable $\w$ increases while keeping a fixed closest impact parameter $y_o$ (rows 1 and 2). In row 3, the lensed signal shows a slightly higher SNR than in the case of the unlensed signal, with the closest impact parameter set at $y_o=1$. This decrease in SNR occurs because the amplification factor diminishes as the closest impact parameter increases, which is evident in the lensed signal. The expected lensing pattern of the lensed signal agrees with analytical predictions when $\w$ increases and $y$ decreases, as shown in Figs.~\ref{fig:snr-segment1} columns 3 and 4 in rows 1 and 2.

For the one-day time series segment, the unlensed signal is below the SNR threshold, and the lensed signal exhibits the same behavior in most cases, as shown in Fig.~\ref{fig:snr-segment1}. Nevertheless, in the case of six-day time series segments the SNR is higher for both the unlensed and lensed signals than the case of one-day time series segment and it is above the SNR threshold. This increase is attributed to the longer segment duration. As mentioned earlier, the SNR is directly proportional to the square root of the segment duration. The trade-off for the lensed signal in the six-day segment is the reduced number of SNR data points, as only the maximum SNR values are considered. The total number of SNR data points for the six-day time series segment analysis is one-sixth of the total number of SNR data points in the one-day time series segment analysis for the same observation period. This reduction in SNR data points causes the expected lensing pattern in the six-day segments time series to appear less definite, as shown in Fig.~\ref{fig:snr-segment6}. 
%----------------------------------------%
% Figure: Comparison
%----------------------------------------%
\begin{figure}[htp!]
   %\raggedleft
    \includegraphics[scale=.42]{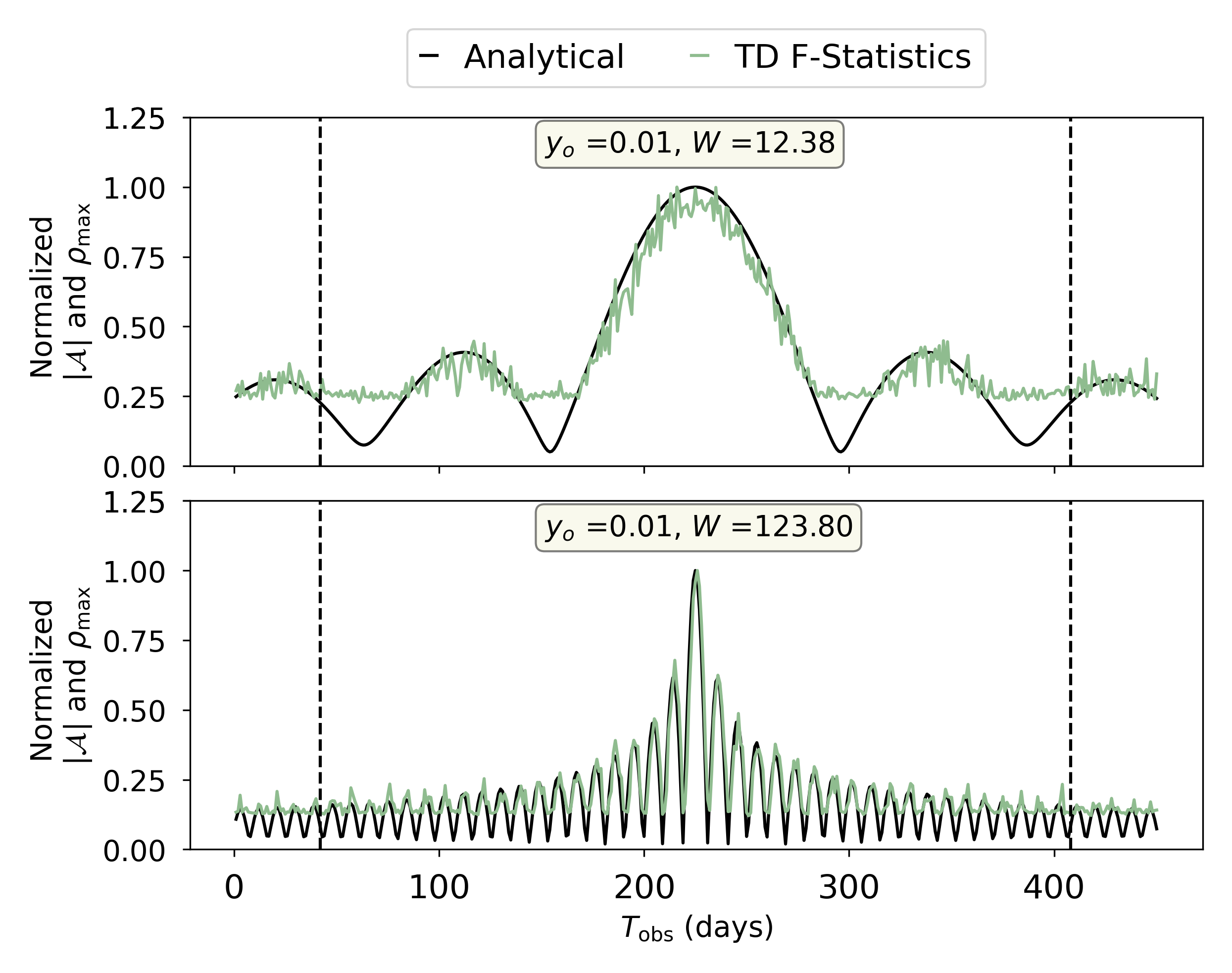}
   \caption{The comparison of lensing patterns between analytical estimation of an amplification factor and the {\tdfstat} results for two example cases. The amplification and SNR from both cases are normalized to 1. 
   Note regions where the signal's amplitude falls below the noise floor, implying that the F-statistic value of the signals is lower than the noise-alone F-statistic value, making it undetectable.}
   \label{fig:pattern-fit}
\end{figure}

The lensing pattern observed from the {\tdfstat} results, particularly at $\w >10$ and $y_o \leq 0.1$, agrees with the expected analytical pattern. The amplification factor from analytical estimations and SNR from {\tdfstat} one-day time series segment analyses are normalized to 1, allowing for a direct comparison of the lensing patterns. Figure~\ref{fig:pattern-fit} shows the two specific patterns $\w=12.38, 123.80$, and $y_o = .01$ which are compared after they are normalized. The predicted lensing pattern from the analytical approach agrees with the results from the time-domain analysis. The flat region near the normalized value of 0.25 (top panel) and 0.2 (bottom panel) in the {\tdfstat} appears because the signal is below the noise floor, making it invisible due to dominance of the noise.

% ================================== %    
% Discussion and conclusions
% ================================== % 
\section{\label{sec:discussion_conclusion} Discussion and conclusions}
Standard CW searches, performed so far on the Advanced Detectors Era data (O1-O3, and O4a, see \cite{LVK-O4a_known_pulsar}) assume that the signal amplitude remains constant throughout the observations. In contrast, lensing causes an evolution of the GW amplitude, with the signal amplitude variations directly impacting the estimation of the SNR, leading to a distinct lensing pattern (the GW ``Paczyński curve''). This, in turn, enhances the detectability of CW signals by making them temporarily ``transient'', i.e., raising their SNR above the noise threshold. In this work, we have studied the projection of the microlensing effect on the CW parameter space. Using simulated examples, we demonstrate that the evolution of the signal-to-noise ratio (SNR) contains information about the amplification factor, which can be recovered using a standard CW search pipeline {\tdfstat} designed for detecting constant amplitude signals.

Specifically, we examined the parameter space associated with the lensing of CW signals by stellar-mass compact lenses of different masses, and studied changes in the amplification over time that result in a lensing amplification pattern. A subset of parameters was selected to simulate a lensed CW signal. Notably, we show that anticipated changes in the SNR across time-domain segments in the {\tdfstat} analysis agree with the theoretically calculated amplification factor, as shown in Fig.~\ref{fig:af-te}; two specific pattern comparisons are also shown in Fig.~\ref{fig:pattern-fit}. The impact of lensing on the signal's SNR is studied through simulations, employing sequences of one-day and six-day time series segments that resemble realistic input data for CW analysis with the {\tdfstat} method. These results are compared with the unlensed CW signals and noise-only results in Sec.~\ref{sec:tdfstat-results}, specifically in Figs.~\ref{fig:snr-segment1} and \ref{fig:snr-segment6}. We also recognize the advantages and disadvantages associated with selecting the time-domain segment length for coherent signal searches in the {\tdfstat} analysis. It is essential to attain a sufficiently high SNR for each individual segment. At the same time, this requirement may lead to the loss of rapid temporal variations in the lensing pattern for specific lensing parameters, as amplitude fluctuations could be ``averaged out'' during a coherent search. A follow-up study is underway to investigate these amplitude changes within the GW data, offering also an opportunity to examine the impact of noise artifacts~\cite{Davis_2021, FAcernese_2023} related to the lensed signal searches. Our study demonstrates, using selected examples, the effect of microlensed CW signals from isolated sources lensed by an isolated object at a constant frequency can be identified with the TDF analysis. The detection probability of lensed signal will be studied further separately. However, additional effects and more complex astrophysical lensing scenarios need further investigation, which we outline below.

\textit{Impact of source's transverse velocity:}
The source's transverse velocity is one factor that affects the Einstein crossing time, but it also impacts the differential Doppler shifts in frequency during lensing~\cite{PhysRevD.80.044009, 2024arXiv241214159S, Zwick_2025}. These frequencies are detectable, depending on the duration of the time series segments in the {\tdfstat} analysis or the lens mass is very high, particularly in strong lensing scenarios. If the Doppler-shifted frequency is smaller than the frequency resolution then this effect becomes negligible.

\textit{Non standard microlensing patterns:} The standard approach assumes a point-like lens and linear relative motion within the lensing system. The lens may consist of a binary or even a multiple system, which is particularly significant when lenses are situated in the central regions of a globular cluster. While this phenomenon has been well studied in EM lensing, there remains a lack of systematic knowledge regarding GW lensing. Additionally, the source itself may also be a binary; numerous binary pulsars are already known~\cite{Manchester_2005}.

\textit{Parallax Effects:} The source takes days or even several weeks to traverse the Einstein radius, and estimating its position in the sky as well as its proper motion is influenced by the Earth's orbital period via parallax effect. This parallax effect introduces asymmetry into the GW lensing pattern~\cite{strong_weak_micro, Mao2008-xo}. We should note that regarding the case we studied here, our justification is twofold. First, the parallax effect is not significant enough to alter our results. Second, accurately accounting for parallax requires knowledge of the source's position. A more comprehensive discussion on the microlensing of compact objects will be addressed in a separate study. It is also worth mentioning that if the source is part of a binary system, its motion around a common center of mass in the source plane introduces a similar, albeit much smaller, effect known as "xallarap"~\cite{strong_weak_micro, Mao2008-xo}.

\textit{Beyond PML model:} Extending the CW lensing studies from isolated stellar-mass objects to incorporate other lens models, including low-mass Dark Matter (DM) halos help understand dark matter properties. Detecting CW lensing signatures induced by low-mass halos (less than $10^{8} M_{\odot}$) will contribute to addressing the current small-scale crisis~\cite{Bullock:2017xww} in observational cosmology. This investigation is presently underway for both second-generation and next-generation detectors and subsequent search methods are being developed. 

\section{\label{sec:data_software-details} Data generation and Software details}

The simulated signal dataset, which includes unlensed, lensed, and noise-only data, together with the corresponding numerical grid and SSB ephemerids files, can be reproduced using the {\tdfstat} package. The details of the input parameters for data generation are discussed in this paper. The microlensing calculations are performed using python~\cite{Python}, with numpy~\cite{harris2020array}, scipy~\cite{2020SciPy-NMeth}, and matplotlib~\cite{Hunter-2007} libraries. The {\tdfstat} analysis is carried out with a {\tdfstat} implementation \cite{Aasi_201408, tdfstat}. 

\section*{\label{sec:acknowledgement} Acknowledgments}
SS would like to thank Mirosław Giersz for the discussions on IMBH and Cluster, Andrzej Królak for his inputs on the Time-Domain F-statistic discussion on the lensing signal, and Andrew Miller for his insights on PBH, Weronika Narloch and Sarang Shashikant Shah discussion on EM-GL data analysis. Michał Małkowski for preliminary studies on parameter estimation methods. SS and SH express gratitude to the Erwin Schrödinger International Institute for Mathematics and Physics (ESI) and the organizers of the workshop ``Lensing and Wave Optics in Strong Gravity'', where part of the long duration strong lensing discussion was carried out with Johan Samsing, Miguel Zumalacárregui, Mikołaj Korzyński and Helena Ubach. The authors would like to thank  Aditya kumar sharma for his comments on it. We acknowledge the support from the Polish National Science Centre through the grants no. 2021/43/B/ST9/01714 and 2023/49/B/ST9/02777, and from the Nicolaus Copernicus Astronomical Center (CAMK PAN) through access to the computing cluster. Authors thank the {\LaTeX} community.
\begin{appendix}
%--------------------------------------------------------%
%--------------------------------------------------------%
\appendix
\section{\label{app:App-A} Lensed CW signal} 

The CW source being in motion relative to both the lens and the observer causes the impact parameter to change over time. Assuming that the CW source emits a monochromatic radiation without any modulation before the lensing occurs, then the amplification factor for the PML model mentioned in Eq.~\ref{eq:pml-af} becomes a time-varying amplitude at a constant frequency $f$.

The Eq.~\ref{eq:pml-af} is expressed in the frequency domain, where amplification factor $\mathcal{A}$ is a function of $\w$ and $y$. Since $\w$ depends on $\ml$ and $f$, we can replace $\w$ with $\ml$ and $f$. Additionally, we represent $y$ with $t$, as $y$ is a function of time. The Eq.~\ref{eq:pml-af} expressed with $f$ and $t$ is denoted by $\mathcal{A}(f, t)$. In the frequency domain the unlensed CW signal is expressed by $\tilde h(f') = A' \cdot \delta{(f'-f)}$ and it represents a sinusoidal signal, where $A'$ is an amplitude of the source signal.
Hence, 
\begin{eqnarray}
    \mathcal{A}(f, t) = \frac{\tilde h^l(f') }{\tilde h(f')} ,
    \label{eq:afration}
\end{eqnarray}
where $\tilde h^l(f')$ is the lensed signal in frequency domain. From the above equation, $\tilde h^l(f') = \tilde h(f') \; \mathcal{A}(f', t)$ .
The inverse Fourier transform of the lensed CW signal is
\begin{eqnarray}
    h^l(t) = \int_{-\infty}^{\infty} df \; \tilde h^l(f') e^{2 \pi i f' t} \;  ,
    \label{eq:hlt}
\end{eqnarray}

Substituting $\tilde h^l(f')$ and then $\tilde h(f')$ into Eq.~\ref{eq:hlt} we have
\begin{eqnarray}
\label{eq:hltexp}
    h^l(t) = A' \int_{-\infty}^{\infty} df \mathcal{A}(f, t) \; \delta{(f'-f)} e^{2 \pi if' t} .   
\end{eqnarray}
Using the property of the delta function, i.e., $\int f(x) \; \delta(x-a) \; dx = f(a)$, we obtain
\begin{eqnarray}
    h^l(t) = A' \mathcal{A}(f, t) \; e^{2\pi if t} ,
    \label{eq:hltdot}
\end{eqnarray}
where $h(t) = A'e^{2\pi ift}$ is the unlensed monochromatic signal in the time-domain. The lensed time domain signal is
\begin{eqnarray}
    h^l(t) = \mathcal{A}(f, t) \; h(t) .
    \label{eq:hltfinal}
\end{eqnarray}
The CW signal model used in this work is based on the CW signal model for NS mentioned in ~\cite{PhysRevD.58.063001}. Considering only the second component $h_{2k}$ of the CW signal and using the four amplitudes $A_{2k}$ we have
\begin{eqnarray}
    h(t) = \sum^{4}_{k=1} A_{2k}\,h_{2k}(t) .
    \label{eq:htns}
\end{eqnarray}
In time-domain, the lensed signal can be understood as the product of the amplification factor and the unlensed signal. It applies to monochromatic signals, assuming no additional modulations or phase shifts in the lensed signal. Once the lensing effect occurs, the signal is detected and appears modulated. Aside from the changes in amplitude or power, the modulation itself remains consistent between the unlensed and lensed signals. The variation in amplitude or power is a direct result of the lensing effect. So, we express our lensed CW signal from NS as
\begin{eqnarray}
    h^l(t) = \mathcal{A}(f,t) \cdot \left( \sum^{4}_{k=1} A_{2k}\,h_{2k}(t) \right) .
    \label{eq:htnslensed}
\end{eqnarray}

\section{\label{app:App-B} Simulation Parameters} 
For the {\tdfstat} simulations, we chose three narrow-banded time series data, with the lower edge of the band frequency at 10 Hz, 100 Hz, and 1000 Hz. The width of the narrow band is set to 1 Hz. The {\tdfstat} implementation works with internal unit of frequency within the band in the $[0,\,\pi]$ range. The injected signal's frequency is set to 1 radian in the internal frequency unit, which corresponds to 0.31830988 Hz; the frequency spin-down $\dot{f}$ is set to zero. The simulated CW signals strain amplitude is set to $h_{\mathrm{o}}=10^{-25}$. The right ascension $\alpha$, and declination  $\delta$, for the unlensed and lensed signals are $266.42 ^{\circ}$ and $-29^{\circ}$, respectively, which points towards the Galactic center. The wobble and polarization angles are both zero. Four lens masses $\ml$ = $10 \; M_\odot, 10^2 \; M_\odot$, $10^3 \; M_\odot$ and $10^4 \; M_\odot$ are selected from the range of $1 \; M_\odot$ to $10^6 \; M_\odot$. These combinations of frequencies and lens masses provide a broader overview for studying the effect of microlensing, and resulting lensing patterns on the CW parameter space with the {\tdfstat} analysis.

Distances, lens mass, and relative velocities influence the Einstein crossing time $\te$. Different combinations of these elements can lead to degeneracies within $\te$. By taking a specific lens mass and setting the duration of the source to cross the Einstein ring to 366 days, it aligns more closely with the O3 observation run, we can interpret plausible astrophysical scenarios across various distances and relative velocities in the considered range. In this study, our primary focus is exploring a microlensing scenario rather than parameter estimation. Consequently, distances and relative velocities will be free parameters for a specified $\te$ and lens mass.

The one-day and six-day time segments are generated for the unlensed, lensed, and noise-alone data, which has a duration of over 450 days. Gaussian noise is generated with a mean of $\mu=0$ and a standard deviation of $\sigma_{\mathrm{n}} = 0.0002$, which approximates the average $\sigma_{\mathrm{n}}$ derived from observations within a specific narrow band frequency. While the noise strain amplitude is typically on the order of $n_{\mathrm{o}}=10^{-20}$, for illustrative purposes, specifically in Fig.~\ref{fig:td_fd_illustration} the noise strain amplitude is taken as $n_{\mathrm{o}}=10^{-23}$.
\end{appendix}
% % ============================================ %
% % ============================================ %
% ------------------------------- %
\bibliography{references}

% =============================== %

\end{document}